\documentclass[sigconf]{acmart}

\usepackage{booktabs} 
\usepackage{wrapfig}

\setcopyright{rightsretained}

\usepackage{booktabs,tabularx}
\usepackage{enumitem}

\author{Manas Satish Bedmutha}
\orcid{0000-0003-3427-2226}
\affiliation{%
\institution{UC San Diego}
  \city{La Jolla}
  \state{California}
  \country{USA}
}
\email{mbedmutha@ucsd.edu}

\author{Tanvi Kandepuneni}
\orcid{0009-0009-8473-8988}
\affiliation{%
  \institution{University of Massachusetts Amherst}
  \city{Amherst}
  \country{USA}
}
\email{tkandepuneni@umass.edu}

\author{Bhawana Chhaglani}
\orcid{0000-0002-4060-4883}
\affiliation{%
  \institution{Penn State University}
  \city{Sate College}
  \country{USA}
}
\email{bbc5655@psu.edu}

\begin{document}
\title{SoundVote: Selecting Test-Time Transformations for Reliable Audio Event Detection}

\begin{abstract}
Pretrained audio foundation models can detect diverse audio events without task-specific fine-tuning, but they suffer from prediction instability and reliability challenges. 
We present \textit{SoundVote}, which combines event labels from selected transformations of the same recording using voting and logistic regression. 
We evaluate pitch shifts, time stretches, gain changes, and time and frequency masking with two audio foundation models (SenseVoiceSmall and Qwen2-Audio) on five human non-speech event classes (breath, cry, cough, sneeze, laugh). 
On ESC-50, nine-view voting improves SenseVoiceSmall accuracy from 63.5\% to 73.0\%, increasing mean processing time from 0.18 to 1.92 seconds per clip. Logistic regression with the same nine views reaches 77.5\% held-out accuracy.
On Qwen2-Audio, nine-view logistic regression improves held-out accuracy from 91.5\% with the original view to 95.5\%. 
Pitch shifting and time stretching were among the most useful transformations as their ensemble improved the accuracy significantly for both models. 
The accuracy gains also extend to a second dataset, Nonspeech7k, where the fixed nine-view ensemble improves SenseVoiceSmall accuracy by 7.4\%. SoundVote improves the reliability of existing audio foundation models while keeping them frozen and gives practitioners a clearer understanding of how transformations affect predictions and more control over how those predictions can be leveraged to improve the accuracy of the target task.


\end{abstract}

\vspace{-0.1cm}

\ccsdesc[500]{Human-centered computing~Ubiquitous and mobile computing}
\keywords{Reliablilty, AudioLLMs, Ensemble modeling, Test-time adaptation, Foundational models}
\maketitle

\section{Introduction}
\label{sec:intro}
Audio foundation models are becoming a common starting point for sound event detection (SED), speech emotion recognition (SER), and other audio understanding tasks \cite{li2024audio, yang2023uniaudio,latif2023sparks, srivastava2026majic}. Released models offer different interfaces for various audio sensing 
tasks. For example, SenseVoiceSmall returns sound-event and emotion tags alongside its transcription~\cite{funaudiollm2024sensevoicesmall}. Qwen2-Audio supports instruction-based audio analysis \cite{chu2024qwen2}, including
sound and emotion recognition. BEATs provides pretrained audio
representations for event classification~\cite{chen2022beats}.
These capabilities make it possible to build generalizable acoustic sensing systems around existing checkpoints. While these models afford newer features, reliance on individual predictions can lead to adverse downstream impact. For example, missed health detections can fail to trigger timely support, incorrect classifications can create contrasting experiences. Model predictions are solely driven by their intrinsic capabilities.


Hence, reliable deployment becomes challenging because of prediction instability i.e. even subtle changes in the input can change the model's decision \cite{zhang2022memo}. Background activity and recording conditions can vary even when the target event is the same. RSA-Bench shows that audio model performance depends on the type and strength of background interference, and TRACE shows that model judgments do not always respond consistently to controlled changes in acoustic evidence~\cite{zhang2026rsa,morais2026trace}. This sensitivity extends to small changes in how a recording is presented. Even minor shift in pitch or amplitude may distort predictions significantly~\cite{zhang2022memo}. Therefore approaches that enhance the robustness of predictions of such black-box models become critical to adoption.

One approach to improving robustness is to expose a classifier to more
acoustic variation during inference. ~\citet{salamon2017deep} show that audio data augmentation can improve environmental sound classification, with effects that vary by sound class. A released model can also be queried on multiple transformed versions of an input at inference time, a technique known as test-time augmentation (TTA). TTA has been well explored in computer vision domain \cite{shanmugam2021better, zhang2022memo} and has also been used in audio-language learning and sound event systems. ~\citet{kim2022exploring}~combine intermediate representations and outputs for audio captioning and audio--text retrieval. Jin Sob Kim et al. rotate multichannel inputs and cluster predictions for sound event localization and detection~\cite{kim2025robust}. Verma and Pilanci use a language model to aggregate traces of sampled patch-level predictions for audio classification, including on ESC-50~\cite{verma2026thinking}. These studies use transformed views, spatial predictions, or patch traces to gather and combine evidence. We study view selection using only the event labels
returned by a frozen audio model without changing the model internals. 

These approaches motivate an inference question, which extra views recover complementary errors, and when are they worth another model call? In image classification, Shanmugam et al. show that a net accuracy gain from TTA can hide changes from correct to incorrect predictions~\cite{shanmugam2021better}. We observe a related trade-off in audio. 
A useful view can therefore have low standalone accuracy and still improve a vote by contributing correct predictions on different clips.

We present \textbf{SoundVote}, a method that combines event labels
predicted by a frozen audio model on different views/transformations of the same audio. SoundVote generates views using pitch shifts, time stretches,
gain changes, and time and frequency masking, then 
combines their predicted labels using voting or a lightweight logistic regression classifier. The core idea is that a transformed view
can be useful even when its standalone accuracy is low. It may correctly
recognize clips that the original view misclassifies, giving the ensemble
a chance to recover those errors. In this work, we measure the errors each view fixes, the correct predictions it overturns, and the inference time it adds.

With SenseVoiceSmall (a foundation model) on ESC-50 (a standard corpus for audio event detection), the original waveform achieves 63.5\%
accuracy at 0.18 seconds per clip. Combining the original with both
pitch-shifted views reaches 70.5\% at 0.68 seconds per clip, and the fixed all nine-view ensemble reaches 73.0\% at 1.92 seconds per clip. 
Using the same nine views, logistic regression reaches 77.5\% held-out accuracy for SenseVoiceSmall and 95.5\% for Qwen2-Audio on ESC-50.
The strength
of a transformation also matters. Increasing the pitch shift from
$\pm1$ to $\pm2$ semitones improves ensemble accuracy, while increasing
it to $\pm3$ changes more predictions in the downward-shifted view
without improving the vote. The best observed combinations vary by
event class. Both time stretches give the highest observed Cough recall
(85.0\%), while combining pitch shifts and a time stretch gives the
highest observed Cry recall (75.0\%) in the exploratory subset search.

Applying the same fixed nine-view ensemble to the five event classes
evaluated on Nonspeech7k dataset~\cite{rashid2023nonspeech7k} raises accuracy from 40.74\% to
48.15\%. The gain on a second dataset shows
that the ensemble can help under different recording conditions.
\noindent
\newline
Through the design of the SoundVote pipeline, we make the following contributions:
\begin{itemize}[leftmargin=*]
  \item We quantify prediction instability in human non-speech event recognition by measuring the impact of small waveform transformations change predictions, recover errors, and overturn correct predictions.
  \item We design the \textit{SoundVote} pipeline, and compare voting and learned aggregation, and characterize how transformation strength and view composition affect recognition accuracy, inference latency, and per-class recall.
  \item We evaluate whether the benefits of combining transformed views
  extend to a second dataset, Nonspeech7k, and a second model,
  Qwen2-Audio.
\end{itemize}

Our findings present that robustness of audio foundation models can be achieved through audio-level augmentations, thereby improving trustworthiness of using large-scale audio models for inference.

\section{Background and Motivation}
\label{sec:motivation}
Pretrained audio models can recognize a range of sound events, but their sensitivity to input variation motivates combining predictions from multiple views of the same recording.

\subsection{Pretrained models for audio event recognition}
Large-scale audio pretraining has produced models that can be reused across audio recognition tasks. PANNs pretrains convolutional audio networks on AudioSet and evaluates their transfer to downstream audio tasks \cite{kong2020panns}. BEATs learns general-purpose audio representations through iterative acoustic tokenization and reports results on audio classification benchmarks  \cite{chen2022beats}. SenseVoiceSmall incorporates an event output into a multilingual speech model and can perform SED and emotion recognition tasks \cite{funaudiollm2024sensevoicesmall,srivastava2025real}. The FunAudioLLM report describes training with pseudo-labeled events in speech data and evaluates event recognition on ESC-50 and other corpora \cite{an2024funaudiollm}. Qwen2-Audio supports instruction-based analysis of speech and other sounds through an audio-language interface \cite{chu2024qwen2}.
The FunAudioLLM benchmark establishes that SenseVoice can recognize the kinds of events we study. We examine a different aspect of that capability by asking how its predictions change across transformed versions of individual clips, which changes help the final decision, and what those additional inference calls cost. Our task is clip-level event recognition, with one reference class per clip.

\subsection{Prediction instability in audio foundation models.}
Audio foundation models are sensitive to small, non-adversarial changes in their input. Adding silence or noise lowers accuracy and increases prediction volatility in audio language models~\cite{li2026silence}. Slight perturbations to multiple-choice audio benchmarks change which answers a model gives~\cite{lopez2025robustness}. Noise, background conversation, and audio quality variation expose further weaknesses across a broad trustworthiness evaluation~\cite{li2026audiotrust}. Pitch, speed, and gain changes degrade speech recognizers~\cite{shah2025speech}, and background sounds that leave human emotion perception unchanged still produce incorrect machine predictions~\cite{jaiswal2021best}. Despite this evidence, existing studies largely characterize instability rather than mitigate it. They quantify accuracy loss or consistency under perturbation, but they do not examine how a practitioner might improve the reliability of a released model that cannot be retrained or inspected. This setting is particularly constrained for frozen, label-only models, where confidence scores and internal representations are unavailable and each additional query increases inference cost. Moreover, the effect of a perturbation on individual predictions is not uniform: a transformed view may recover some errors while overturning correct predictions, so aggregate accuracy alone obscures its utility. We address this gap by treating prediction instability as a source of complementary evidence. We combine the labels returned by a frozen model on transformed views of the same audio, and we measure which views recover errors, which overturn correct predictions, and what inference cost they incur.

\subsection{Audio augmentation during training and inference}
Audio augmentation is widely used in training to vary the acoustic evidence presented to a model. Salamon and Bello found that training-time transformations affect environmental sound classes differently, motivating class-conditional augmentation \cite{salamon2017deep}. SpecAugment introduced time and frequency masking as a training procedure for automatic speech recognition \cite{park2019specaugment}. These studies motivate our choice to examine transformations by class and to inspect their effects on individual predictions.

Audio systems have also combined alternate views at inference time. For example, Xiao et al.'s DCASE 2023 submission averages predictions from original and energy-weighted features for sound event detection \cite{xiao2023fmsg}. Our study compares multiple families of transformations and evaluates their effects on a released model's output labels. It asks which views bring complementary predictions and measures the cost of obtaining them.

\subsection{Aggregation and selection at test time}
Test-time augmentation combines predictions from transformed versions of a test input. In the computer vision domain, Shanmugam et al. demonstrated that averaging predictions can improve aggregate accuracy while still introducing errors on previously correct examples, and proposed a learned aggregation method \cite{shanmugam2021better}. 
Another related line of work adapts pretrained vision models at test time and studies the problem of test time robustification. MEMO uses multiple augmented views to update model parameters by minimizing marginal prediction entropy \cite{zhang2022memo}. 
We apply this correction-and-corruption perspective to non-speech audio events and evaluate view combinations under different inference budgets. 

In audio domain, CALM improves few-shot classification with frozen audio-language models by using reliability-weighted votes from internal attention-head representations \cite{mehta2026calm}. Our experiments keep model parameters fixed during inference and combine the event labels returned for transformed waveforms. This setup lets us study view selection using the model's exposed predictions.
\section{\textit{SoundVote} Design and implementation}
\label{sec:system}

In this section, we describe the design and implementation of the proposed system.
\begin{figure*}
    \centering
    \includegraphics[width=0.8\linewidth]{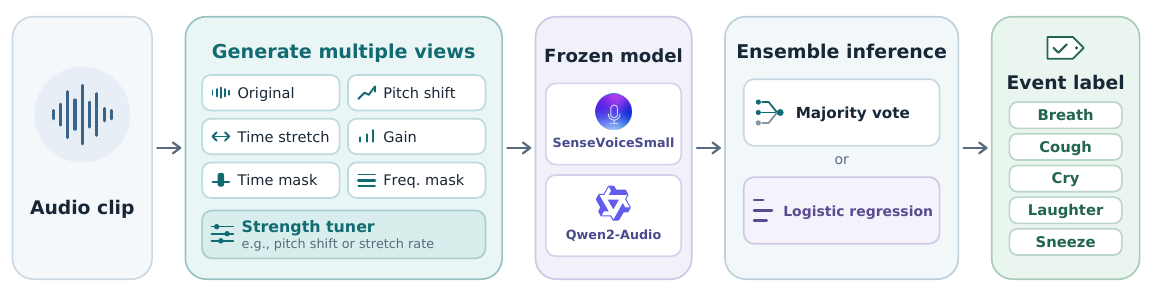}
    \caption{SoundVote design: Test-time ensemble inference for audio event detection. Candidate views of each audio clip include the original waveform, pitch shifts, time stretches, gain changes, and time or frequency masking. A single frozen model, such as SenseVoiceSmall or Qwen2-Audio, predicts an event for each selected view. The ensemble combines these predictions into one clip-level label using policy-based and logistic regression-based approach.}
    \label{fig:system}
\end{figure*}

\subsection{Generating audio views}
SoundVote takes a single audio clip and a policy specifying which views to query. The candidate pool includes the original waveform and eight deterministic transformations. As shown in Table \ref{tab:aug}, pitch shifts move spectral content, time stretching changes how an event unfolds, gain adjusts waveform amplitude, and masking removes a portion of the signal. We use these transformations as alternate presentations of the same reference event. 
All these transformations preserve the audio event in the audio file.
The \textit{strength tuner} controls the magnitude of each transformation, such
as the pitch shift in semitones, the time-stretch rate, the gain in
decibels, or the fraction of the signal masked. In our strength
experiment, we vary the pitch shift and time-stretch rate across
configurations. 


\begin{table}[t]
  \centering
  \small
  \caption{Candidate test-time transformations in SoundVote.}
  \label{tab:soundvote-transformations}
  \small
  \setlength{\tabcolsep}{3pt}
  \renewcommand{\arraystretch}{1.12}
  \begin{tabularx}{\columnwidth}{@{}
    >{\raggedright\arraybackslash}p{0.19\columnwidth}
    >{\raggedright\arraybackslash}p{0.22\columnwidth}
    >{\raggedright\arraybackslash}X
    @{}}
    \toprule
    \textbf{Family} & \textbf{Candidate views} & \textbf{Implementation} \\
    \midrule
    Original & Identity &
      Use the original waveform. \\
    Pitch & $-2$ and $+2$ semitones &
      Shift pitch with \texttt{librosa}. \\
    Time stretch & Rates 0.9 and 1.1 &
      Slow down or speed up the waveform with \texttt{librosa}. \\
    Gain & $-6$ and $+6$\,dB &
      Multiply samples by $10^{g/20}$ for gain $g$ in dB.
      Rescale to unit peak if the peak magnitude exceeds 1. \\
    Time mask & Central mask &
      Zero the middle 10\% of waveform samples. \\
    Frequency mask & Central frequency band &
      Zero the middle 10\% of STFT frequency bins in every frame
      (1{,}024-point FFT, 256-sample hop). Invert to the original length. \\
    \bottomrule
  \end{tabularx}
  \label{tab:aug}
  \vspace{-0.2cm}
\end{table}
\normalsize

\subsection{Per-view Model Inference}
The same frozen model processes each selected view independently. For SenseVoiceSmall, the parser extracts the first recognized event in a fixed tag order from the model's rich transcription and records an unknown or missing event when no recognized event is available. 
For Qwen2-Audio-7B-Instruct \cite{chu2024qwen2}, we pair the audio with the same prompt for the original clip and every transformed view. The instruction reads “Listen to the audio and classify the main sound. Choose exactly one label: Breath, Cough, Cry, Laughter, or Sneeze. Breath means a sigh. Reply with the exact label only.” This is because we observed that Qwen predicts sigh for breathing events. The five labels/categories are selected as the largest comprehensive subset of non-speech human sounds.

\subsection{Ensemble Methods}

\noindent\textbf{Majority vote}. We aggregate the resulting categorical outputs by plurality vote. Unknown and missing events are abstained. If every view abstains, the ensemble returns no event. Other recognized event labels, such as speech, remain eligible to vote. A final label outside the five reference classes counts as a classification error.
We use two tie conventions in the reported SenseVoiceSmall experiments. The fixed-ensemble comparison resolves a tie by the first matching label in the recorded view order, with the original view listed first. The subset search explicitly prefers the original prediction if it is tied for the most votes, then uses lexicographic order among the remaining tied labels. We report accuracy comparisons within each convention.

\noindent\textbf{Selecting views under a latency budget}.
For SenseVoiceSmall, the runtime policy applies a fixed set of views to every clip. To compare candidate policies, we cache the model's predictions and measured costs for each view. We enumerate all 256 subsets that include the original and any combination of the eight transformations. For each subset, we aggregate the cached labels, calculate clip accuracy, and estimate sequential latency by summing its per-view costs. Sorting the subsets by latency and retaining candidates that improve accuracy yields an empirical accuracy–latency frontier. This exposes configurations for different inference budgets and accuracy tolerance.

\noindent\textbf{Learned aggregation with logistic regression.}
SoundVote uses multinomial logistic regression to combine the
labels returned by frozen SenseVoice and Qwen models for the
original audio and eight transformed views. SenseVoice labels
are extracted from event tags, while Qwen labels are parsed
from generated responses. We independently one-hot encode each
view's prediction and concatenate the encodings, retaining
additional outputs such as \texttt{Event\_UNK} and \texttt{Speech}.
For Qwen, separate classifiers
learn class-specific weights for each view--label combination
and an intercept, returning the highest-scoring class.
We use L2 regularization ($C=1$), L-BFGS, and at most 2,000
iterations, without hyperparameter tuning. Evaluation uses
ESC-50's five predefined folds: 160 training and 40 test clips
per round, with source recordings kept together and encoders
fitted only on training folds. Both models remain frozen;
classifier training uses cached predictions.


\section{Dataset and Experimental Setup}
We evaluate SoundVote on five human non-speech event classes, measuring how test-time transformations affect recognition accuracy and inference latency across models and datasets.
\newline
\newline
\noindent\textit{Dataset and task}. 
We select two datasets (ESC-50 and NonSpeech7k) for detecting audio events, on which SenseVoice and Qwen models are not trained. ESC-50 contains five-second audio clips organized into five folds that keep clips from the same source recording together~\cite{piczak2015esc}. We evaluate all 200 clips from five human non-speech event classes: breathing, coughing, baby crying, laughter, and sneezing. Each class contains 40 clips.
Nonspeech7k contains 7,014 mono recordings sampled at 32\,kHz, with durations from 0.5 to 4\,s~\cite{rashid2023nonspeech7k}. We use the official test split, which contains 725 clips across seven classes. Of these, 605 clips belong to the five event categories evaluated on ESC-50. After excluding clips whose source identifiers matched those in ESC-50, we retain 513 clips. Nonspeech7k labels crying as a general category, whereas ESC-50 uses baby crying.


\noindent
\newline
\textit{Models implementation}. We use the released FunAudioLLM's SenseVoiceSmall checkpoint through FunASR 1.4.16 in Google Colab. We load each recording as mono at its native sample rate and construct views with NumPy and librosa. After augmentation, we apply peak scaling when the absolute waveform peak exceeds 1. We write each view to a temporary WAV at the source sample rate and call the model with automatic language identification, inverse text normalization disabled, and a fresh cache. The saved augmentation run contains 1,800 successful clip–view evaluations. The original-view baseline uses the same WAV-based inference and single-label parsing path as the augmented views.
We use the released Qwen2-Audio-7B-Instruct checkpoint~\cite{chu2024qwen2}.
We resample each audio view to 16\,kHz and pair it with the same
classification prompt, which asks the model to return one of five labels:
Breath, Cough, Cry, Laughter, or Sneeze. We map \texttt{Sigh} to
\texttt{Breath} before scoring and ensemble voting.
\noindent
\newline
\textit{Baselines and protocol}. We compare the original view, each transformation alone, family-specific three-view ensembles, a five-view pitch-and-stretch ensemble, and the full nine-view ensemble. We then evaluate the exhaustive original-containing subset search to find the optimal subset. SenseVoiceSmall subset selection and scoring use the same 200 clips. We report selected subsets as exploratory results and use the fixed configurations to characterize the benchmark trade-off. 
\noindent
\newline
\textit{Metrics and timing}. We report clip accuracy and per-class recall. A correction occurs when a transformed view gets a clip right that the original gets wrong. A corruption occurs when the transformed view gets a clip wrong that the original gets right. Both rates use all evaluated clips as the denominator. We time waveform transformation and the model call separately, synchronizing CUDA around the model call. Ensemble latency is the mean of the sum of these costs across a clip's selected views. It includes the model call's processing of the temporary WAV and excludes the separate source-file load, temporary WAV write, and vote. 
We ran inference on an NVIDIA Tesla
T4 GPU and we processed one clip at a time and
evaluated ensemble views sequentially. The reported latency therefore describes measured processing under sequential execution in seconds.
\section{Evaluation}
In this section, we first quantify how small waveform transformations change event predictions, then examine which views to combine, how to aggregate them, and whether the resulting accuracy gains justify the added latency and carry over to another model and dataset.

\noindent
\newline
\textbf{Original-view baseline}. On the original audio, SenseVoiceSmall correctly classified 127 of 200 clips (aggregate accuracy: 63.5\%). Per-class recall was 87.5\% for Breath, 80.0\% for Cough, 47.5\% for Cry, 75.0\% for Laughter, and 27.5\% for Sneeze. Cry and Sneeze accounted for 50 of the 73 baseline errors.

\subsection{Prediction Changes and Error Recovery by Transformation}

Table \ref{tab:single_view} shows the single-view aggregate accuracy (on 5 classes) of all the transformations. The Gain increase and frequency masking each achieve 65.0\% accuracy, 1.5 percentage points above the original. They recover three original errors each and introduce no new errors in this run. Pitch and stretch perturbations change predictions on 25.5–30.0\% of clips, substantially more often than gain increase or frequency masking. The +2 semitone view, for example, fixes 18 original mistakes but breaks 29 correct original predictions. Its standalone accuracy consequently drops to 58.0\%. This combination of recoveries and new errors makes it a candidate for voting even when replacing the original with it would hurt performance.

\begin{table}[t]
  \centering
  \caption{SenseVoiceSmall single-view accuracy and changes from the original
  prediction. Fixed and broken are clip counts.}
  \label{tab:soundvote-single-view}
  \small
  \setlength{\tabcolsep}{3pt}
  \renewcommand{\arraystretch}{1.12}
  \begin{tabularx}{\columnwidth}{@{}>{\raggedright\arraybackslash}Xrrrr@{}}
    \toprule
    \textbf{View}
    & \shortstack[r]{\textbf{Accuracy}\\(\%)}
    & \shortstack[r]{\textbf{Changed}\\(\%)}
    & \textbf{Fixed}
    & \textbf{Broken} \\
    \midrule
    Original                  & 63.5 &  0.0 &  0 &  0 \\
    Pitch $-2$ semitones      & 55.0 & 28.5 & 14 & 31 \\
    Pitch $+2$ semitones      & 58.0 & 30.0 & 18 & 29 \\
    Stretch 0.9               & 53.5 & 25.5 & 10 & 30 \\
    Stretch 1.1               & 56.0 & 28.0 & 14 & 29 \\
    Gain $-6$\,dB             & 61.0 &  5.5 &  2 &  7 \\
    Gain $+6$\,dB             & 65.0 &  1.5 &  3 &  0 \\
    Time mask                 & 54.5 & 15.0 &  4 & 22 \\
    Frequency mask            & 65.0 &  2.5 &  3 &  0 \\
    \bottomrule
  \end{tabularx}
  \label{tab:single_view}
\end{table}

\subsection{Ensemble Accuracy and Latency}

\begin{figure}
    \centering
    \includegraphics[width=\linewidth]{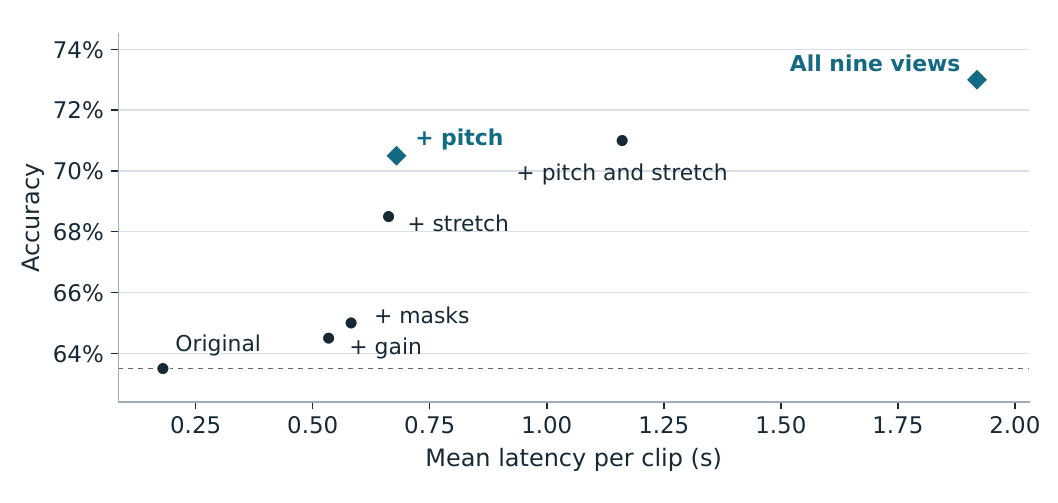}
    \caption{Accuracy and latency of fixed SenseVoiceSmall ensembles. Each point shows the accuracy and mean sequential latency of one configuration. A plus sign indicates views added to the original waveform. The dashed line marks the 63.5\% original-view baseline.}
    \label{fig:subset}
\end{figure}

Figure \ref{fig:subset} shows the results of combining different transformations with original view and their majority vote results. 
The fixed three-view pitch ensemble (original, +pitch, -pitch) improves accuracy by 7.0 percentage points even though both pitch views are individually were weaker than the original. It also outperforms the other three-view families. The fixed nine-view ensemble improves accuracy by 9.5 points, reaching 146 correct clips compared with 127 for the original. That improvement raises mean sequential latency from 0.180 to 1.919 seconds per clip. The three-view pitch ensemble achieves 70.5\% in 0.679 seconds per clip, making it suitable from accuracy-latency tradeoff perspective as it captures about three-quarters of the observed accuracy gain at about one-third of the total measured latency.


\noindent
\newline
\textbf{Class-wise analysis}. 
Table \ref{tab:class_wise} shows the class-level recall for the original view and the fixed
nine-view SenseVoiceSmall ensemble. The largest recall gain comes from baby cry, which improves from 47.5\% to 70.0\%. Sneeze recall improves from 27.5\% to 40.0\%. These two classes account for 14 of the 19 net additional correct labels. Sneeze remains the weakest class even after aggregation.

\begin{table}[t]
  \centering
  \caption{Per-class recall for the original view and the fixed nine-view
  SenseVoiceSmall ensemble. Changes are in percentage points (pp).}
  \label{tab:soundvote-class-recall}
  \small
  \setlength{\tabcolsep}{3pt}
  \renewcommand{\arraystretch}{1.12}
  \begin{tabularx}{\columnwidth}{@{}>{\raggedright\arraybackslash}Xrrr@{}}
    \toprule
    \textbf{Event}
    & \shortstack[r]{\textbf{Original}\\(\%)}
    & \shortstack[r]{\textbf{All views}\\(\%)}
    & \shortstack[r]{\textbf{Change}\\(pp)} \\
    \midrule
    Breath   & 87.5 & 92.5 & $+5.0$  \\
    Cough    & 80.0 & 82.5 & $+2.5$  \\
    Cry      & 47.5 & 70.0 & $+22.5$ \\
    Laughter & 75.0 & 80.0 & $+5.0$  \\
    Sneeze   & 27.5 & 40.0 & $+12.5$ \\
    \bottomrule
  \end{tabularx}
  \label{tab:class_wise}
\end{table}

\noindent
\newline
\textbf{Subset selection}. Table \ref{tab:subset_serch} shows the optimal subset under different latency budgets. Under the subset-search tie convention, the fastest configuration to reach the search maximum of 72.5\% uses the original, both pitch shifts, the 1.1 stretch, and the time mask. It is similar to the the nine-view configuration under this convention while reducing measured latency by approximately 43\%. A three-view candidate using the original, the +2 semitone shift, and the 1.1 stretch reaches 71.5\% in 0.672 seconds. The subset search identifies useful candidates, with post-selection accuracy measured on the same clips used to choose the subsets. 

\begin{table}[t]
  \centering
  \caption{Exploratory SenseVoiceSmall subset search. Subsets were selected
  and scored on the same clips under the subset-search tie rule.}
  \label{tab:soundvote-subset-search}
  \small
  \setlength{\tabcolsep}{3pt}
  \renewcommand{\arraystretch}{1.12}
  \begin{tabularx}{\columnwidth}{@{}>{\raggedright\arraybackslash}Xrrr@{}}
    \toprule
    \textbf{Subset}
    & \textbf{Views}
    & \shortstack[r]{\textbf{Accuracy}\\(\%)}
    & \shortstack[r]{\textbf{Latency}\\(s/clip)} \\
    \midrule
    Original + pitch $-2$                              & 2 & 68.5 & 0.423 \\
    Original + pitch $+2$ + stretch 1.1                 & 3 & 71.5 & 0.672 \\
    Original + both pitches + stretch 1.1 + time mask  & 5 & 72.5 & 1.094 \\
    All views                                          & 9 & 73 & 1.919 \\
    \bottomrule
  \end{tabularx}
  \label{tab:subset_serch}
\end{table}

\subsection{Learned Aggregation with Logistic Regression}
\label{sec:eval-learned-aggregation}

Table \ref{tab:LR} shows the results of using Logistic regression as the ensemble inference strategy. It correctly classifies 155 of 200 ESC-50 clips
(77.5\%) using the labels from all nine SenseVoiceSmall views, resulting in 4.5\% accuracy improvement over the majority voting ensemble scheme. The major1 gain in accuracy is coming from Sneeze accuracy, which increased to 60\% using LR.

\begin{table}[t]
  \centering
  \small
  \caption{SenseVoiceSmall logistic regression results on ESC-50.
  Each ablation drops one transformed view. Changes are measured in
  percentage points relative to nine-view logistic regression.}
  \label{tab:learned-aggregation-results}
  \begin{tabular}{@{}lrrrr@{}}
    \toprule
    Configuration & Views & Correct & Accuracy & $\Delta$ (pp) \\
    \midrule
    Logistic regression & 9 & 155 & 77.5\% & --- \\
    \midrule
    \multicolumn{5}{@{}l}{\textit{View removed}} \\
    Pitch $-2$ semitones & 8 & 153 & 76.5\% & $-1.0$ \\
    Pitch $+2$ semitones & 8 & 148 & 74.0\% & $-3.5$ \\
    Time stretch $0.9$ & 8 & 155 & 77.5\% & 0.0 \\
    Time stretch $1.1$ & 8 & 156 & 78.0\% & $+0.5$ \\
    Gain $-6$ dB & 8 & 155 & 77.5\% & 0.0 \\
    Gain $+6$ dB & 8 & 151 & 75.5\% & $-2.0$ \\
    Time mask & 8 & 154 & 77.0\% & $-0.5$ \\
    Frequency mask & 8 & 151 & 75.5\% & $-2.0$ \\
    \bottomrule
  \end{tabular}
  \label{tab:LR}
  \normalsize
\end{table}
To
measure each transformation's contribution to this ensemble, we omit
one transformed view at a time, keep the original view, and refit
logistic regression on the remaining views using the same five
held-out folds.
Removing the $+2$ semitone pitch shift produces the largest drop,
from 77.5\% to 74.0\% (seven fewer correct clips). Removing either
the frequency mask or the $+6$ dB gain view lowers accuracy by 2.0
percentage points. Removing the $1.1$ time stretch yields one
additional correct prediction, raising accuracy to 78.0\%.

\subsection{Class-Specific Transformation Selection}

Different sound events may benefit from different transformations. For a
targeted detector, such as a cough detector, this raises a practical question:
which views recover the most examples of the target event? We search all 256
ensembles that contain the original waveform and any subset of the eight
transformed views. For each event class, we select the subset with the highest
recall, breaking ties in favor of lower mean latency. This search gives the
best observed recall within the tested combinations. This can be useful for developers to identify the best transformations for their target task and latency budgets. 
Table \ref{tab:class_specific_transformations} shows the best performing ensemble set for each class. 
\begin{table}[t]
  \centering
  \caption{Best in-sample SenseVoiceSmall ensemble for each event class (majority voting).
  Every ensemble includes the original waveform. Gain is the change
  in recall in percentage points (pp).}
  \label{tab:class-specific-ensembles}
  \small
  \setlength{\tabcolsep}{3pt}
  \renewcommand{\arraystretch}{1.10}
  \begin{tabularx}{\columnwidth}{@{}lrrr>{\raggedright\arraybackslash}X@{}}
    \toprule
    \textbf{Class}
    & \shortstack[r]{\textbf{Original}\\(\%)}
    & \shortstack[r]{\textbf{Best}\\(\%)}
    & \shortstack[r]{\textbf{Gain}\\(pp)}
    & \textbf{Views added to the original} \\
    \midrule
    Breath   & 87.5 & 92.5 & $+5.0$  & Time mask \\
    Cough    & 80.0 & 85.0 & $+5.0$  & Stretches 0.9 and 1.1 \\
    Cry      & 47.5 & 75.0 & $+27.5$ & Pitch $\pm 2$, stretch 1.1 \\
    Laughter & 75.0 & 80.0 & $+5.0$  & Stretch 1.1, gain $+6$\,dB \\
    Sneeze   & 27.5 & 40 & $+12.5$ & Pitch $+2$, stretch 1.1 \\
    \bottomrule
  \end{tabularx}
  \label{tab:class_specific_transformations}
\end{table}
\paragraph{The best combination varies by event.}
For Breath, adding only the central time mask increases recall from
87.5\% to 92.5\%. For Cough, combining the original with both time stretches
increases recall from 80.0\% to 85.0\%. Cry shows the largest gain, from
47.5\% to 75.0\%, when we add both pitch shifts and the 1.1 stretch.
Laughter improves from 75.0\% to 80.0\% with the 1.1 stretch and $+6$\,dB
gain. Sneeze improves from 27.5\% to 40\% with the $+2$-semitone pitch
shift and the 1.1 stretch. For a cough detector, these results make the
three-view stretch ensemble a candidate for further evaluation using both
recall and precision.

\paragraph{The selected views suggest hypotheses about acoustic cues.}
For breathing, masking the center of the clip may suppress a brief
distractor while leaving enough airflow sound elsewhere for recognition.
For cough, the two stretches may help by presenting the same burst at
different durations. For crying, pitch shifts and stretching may provide
complementary views of the pitch contour and its timing. For laughter,
stretching may alter the spacing of repeated bursts, while increasing gain
may help the model recognize low-amplitude examples. For sneezing, the
pitch and stretch views may change how the model responds to the spectrum
and duration of a brief burst. These are hypotheses suggested by the
selected subsets. We further evaluate the performance on a second dataset to verify these findings (See Section \ref{sec:nonSpeech_dataset}).

\paragraph{Some classes have more room for improvement than others.}
The best Cry subset recognizes 30 of 40 clips, close to the 32 clips for
which at least one view is correct. The best Sneeze subset recognizes
15 of 40 clips, while at least one view is correct on 22. That seven-clip
gap shows that the tested views contain useful predictions that a single
class-specific voting subset leaves unrecovered.

\subsection{Prediction Disagreement as an Error Signal}
\begin{figure}
    \centering
    \includegraphics[width=0.6\linewidth]{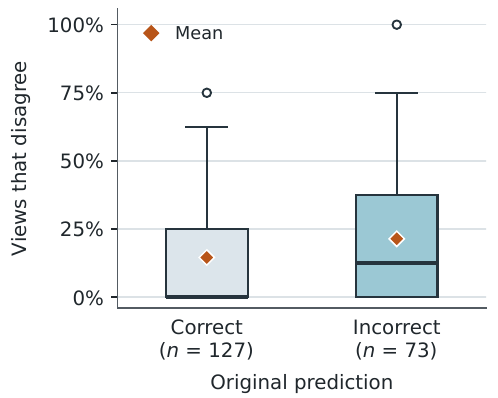}
    \caption{Prediction instability for correct and incorrect original SenseVoiceSmall predictions on ESC-50. For each clip, instability is the fraction of the eight transformed views that disagree with the original prediction. Diamonds indicate group means.}
    \label{fig:instability}
\end{figure}

Prior work has used disagreement as a error signal for hallucinations in generative models \cite{chhaglani2026artifree}. 
We define instability as the fraction of the eight
transformed views whose predictions differ from the original-view prediction.
On average, transformed views disagree with incorrect original predictions
21.4\% of the time and with correct original predictions 14.6\% of the time as shown in Figure \ref{fig:instability}.
The difference is 6.8 percentage points (95\% clip-level bootstrap interval
[0.7, 13.3] percentage points, 10{,}000 resamples). An exploratory one-sided
Mann--Whitney test gives $U=5551.5$ and $p=0.0064$.
Disagreement therefore provides a modest signal for identifying clips that may warrant another inference pass.
The two groups still overlap substantially. Instability identifies original
errors with an AUROC of 0.599 and an average precision of 0.428, compared
with an overall error prevalence of 0.365. We also test a single extra view
as a lower-cost probe. When the $+2$-semitone view disagrees with the original,
the original prediction is wrong on 51.7\% of clips, compared with 30.0\%
when the views agree. Flagging disagreements catches 31 of the 73 original
errors and adds 0.256 seconds per clip on average. Disagreement can therefore
help prioritize clips for review. It has some value for detecting hallucinations/errors \cite{chhaglani2026clear, chhaglani2026artifree} and deciding when to run more
views against the added inference cost.


\subsection{Prediction-Based View Selection}
\label{subsec:prediction-routing}

\paragraph{How routing works.}
A fixed ensemble runs the same views for every clip. The router first
runs SenseVoiceSmall on the original waveform and uses the returned label
to look up which views to run next. If the selected subset contains only
the original view, the router returns that prediction immediately.
Otherwise, it runs the selected transformations and votes over their
labels together with the original prediction, reusing the first model call.
\paragraph{Selecting the views.}
On the development folds, we group clips by the label predicted from
the original waveform. For each group, we evaluate all 256 subsets that
include the original view and select the subset that correctly classifies
the most clips in that group. We break ties by choosing the subset with
lower mean processing time. For labels with fewer than five development
clips, we use the best subset across all development clips. We use that
same overall subset as the fixed policy.
\paragraph{Held-out results.}
We select the views for both policies on four source-grouped ESC-50 folds
and evaluate them on the remaining fold. We rotate through all five folds
and pool the held-out predictions. Both policies break voting ties in favor
of the original label when it is among the tied labels, and otherwise use
lexicographic order. Processing time includes the original model call.
As shown in Table~\ref{tab:sensevoice-routing}, the router achieves
71.5\% accuracy, compared with 70.5\% for the fixed subset and 63.5\%
for the original view alone. Both policies correct 16 original errors.
The fixed subset introduces two new errors, while the router preserves
all originally correct predictions in this evaluation. The difference
in accuracy is two clips. The larger difference is processing time,
which drops from 1.093 to 0.477 seconds per clip, a 56\% reduction
relative to the fixed subset.

The learned view choices explain the reduction. Across all five folds,
an original prediction of Breath, Cough, Cry, or Sneeze triggered no
extra views. An unknown event or Speech prediction always triggered
extra views, and Laughter triggered extra views in four of the five
folds. The router therefore spends additional inference time on clips
with particular initial predictions.

\begin{table}[t]
  \centering
  \caption{Held-out ESC-50 policy comparison for SenseVoiceSmall.
  Fixed and routed subsets are selected on four folds and evaluated on
  the fifth. Fixes and breaks are counts relative to the original view.}
  \label{tab:sensevoice-routing}
  \small
  \setlength{\tabcolsep}{3pt}
  \renewcommand{\arraystretch}{1.12}
  \begin{tabularx}{\columnwidth}{@{}>{\raggedright\arraybackslash}Xrrrr@{}}
    \toprule
    \textbf{Policy}
      & \shortstack[r]{\textbf{Accuracy}\\(\%)}
      & \textbf{Fixed}
      & \textbf{Broken}
      & \shortstack[r]{\textbf{Latency}\\(s/clip)} \\
    \midrule
    Original only       & 63.5 &  0 & 0 & 0.180 \\
    Fixed subset        & 70.5 & 16 & 2 & 1.093 \\
    Prediction-based router & 71.5 & 16 & 0 & 0.477 \\
    \bottomrule
  \end{tabularx}
\end{table}

\subsection{Effect of transformation strength}
We vary pitch shifts across ±1, ±2, and ±3 semitones and compare three pairs of stretch rates, 0.95/1.05, 0.90/1.10, and 0.85/1.15. Each setting votes over the original clip and the two transformed views using the fixed-ensemble rule. This keeps the number of model calls constant across settings. We also measure the fraction of clips for which each individual transformation changes the original prediction.
\begin{figure}
    \centering
    \includegraphics[width=\linewidth]{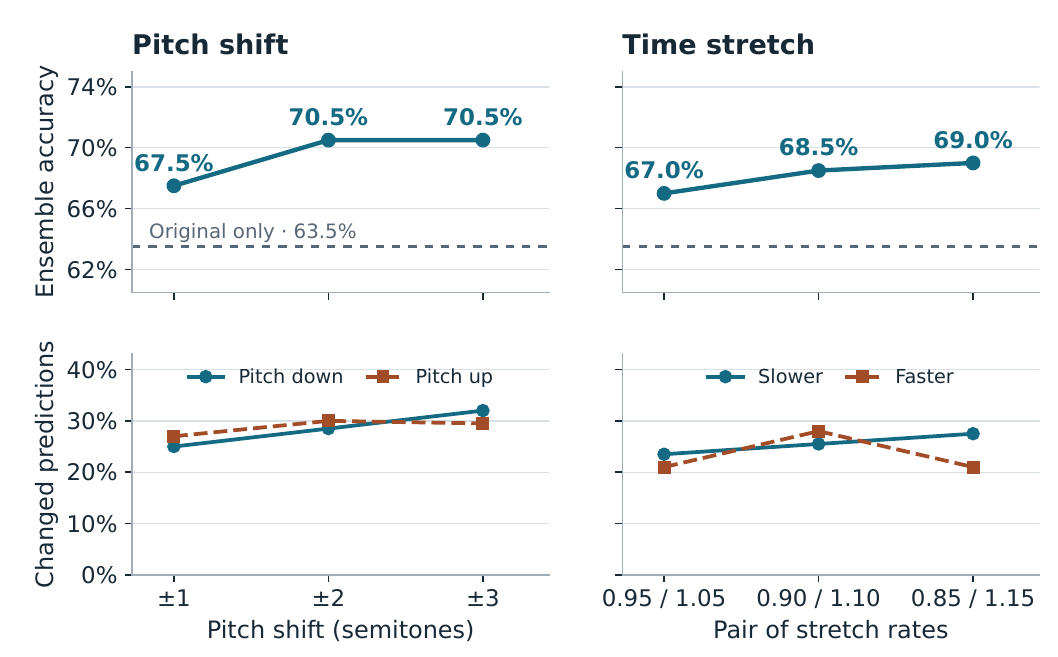}
    \caption{Effect of transformation strength on SenseVoiceSmall. The top panels show accuracy for three-view ensembles that combine the original clip with a pair of pitch or stretch views. The bottom panels show how often each individual transformed view changes the original prediction.}
    \label{fig:acc_latency}
    \vspace{-0.3cm}
\end{figure}

Figure \ref{fig:acc_latency} shows the accuracy variation with the strength of transformation.
Increasing the pitch shift from ±1 to ±2 semitones raises ensemble accuracy from 67.5\% to 70.5\%. Increasing the shift to ±3 leaves accuracy at 70.5\%, even as the pitch-down view's disagreement rate rises from 28.5\% to 32.0\%. For time stretching, accuracy increases from 67.0\% for the narrowest rate pair to 68.5\% for 0.90/1.10 and 69.0\% for 0.85/1.15. That final increase is one additional correct clip out of 200. For the faster view alone, the rate of changed predictions rises from 21.0\% at 1.05 to 28.0\% at 1.10, then drops to 21.0\% at 1.15.

These results show that a larger transformation can produce different predictions even after ensemble accuracy has plateaued. Among the tested settings, \textit{±2 semitones is the smallest pitch shift to reach the maximum observed pitch-ensemble accuracy.} The strength comparison uses the same ESC-50 clips for all settings, so its rankings are descriptive.

\subsection{Cross-Dataset Evaluation on Nonspeech7k}
\label{sec:nonSpeech_dataset}
\begin{figure}
    \centering
    \includegraphics[width=\linewidth]{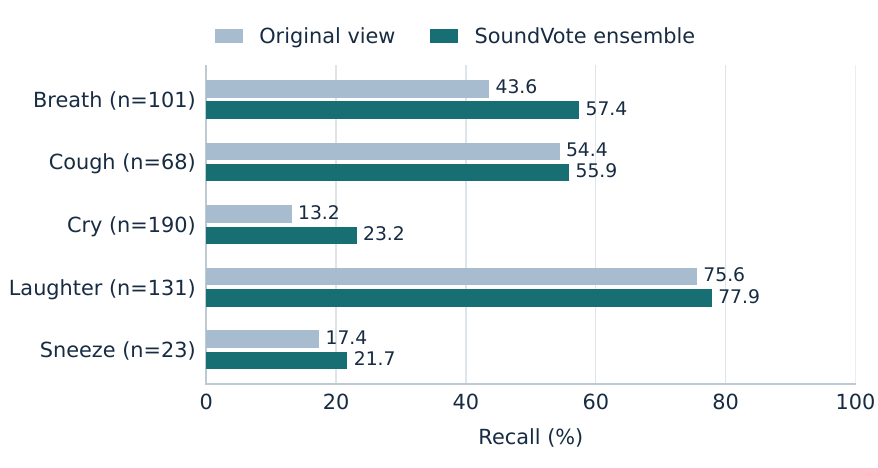}
    \caption{Cross-Dataset Evaluation on Nonspeech7k using SoundVote}
    \label{fig:nonspeech}
    \vspace{-0.5cm}
\end{figure}

We evaluated SoundVote with SenseVoiceSmall on the same five event categories in ESC-50 and NonSpeech7k: breathing, coughing, crying, laughter, and sneezing. Figure \ref{fig:nonspeech}  shows the results before and after ensemble the datasets. 
On the 513 NonSpeech7k clips, accuracy increased from 40.74\% to 48.15\%, yielding 38 additional correct predictions and a gain of 7.41 percentage points. The ensemble therefore improved accuracy on both datasets despite the substantially lower original-view accuracy on NonSpeech7k.

On NonSpeech7k, recall improved in all five classes. The largest gains were for breathing, from 43.6\% to 57.4\%, and crying, from 13.2\% to 23.2\%. Together, these two classes accounted for 33 of the 38 net additional correct predictions. The gains also extended to sneezing (17.4\% to 21.7\%), laughter (75.6\% to 77.9\%), and coughing (54.4\% to 55.9\%). Laughter had the highest original-view recall and still benefited from aggregation, while crying remained the lowest-recall class despite its 10-point gain.

Across the two datasets, the ensemble’s accuracy advantage persisted under different baseline conditions. The NonSpeech7k results show that this advantage extended across all evaluated event categories, with most of the additional correct predictions coming from breathing and crying. Despite consistent gains across datasets, ensemble accuracy on NonSpeech7k remained below 50\%. This motivates further work on improving models and cross-dataset robustness, particularly for crying and sneezing, where the ensemble still missed more than three-quarters of clips.
\vspace{-0.2cm}
\subsection{Cross-Model Evaluation on Qwen2-Audio}
We evaluate SoundVote with frozen Qwen2-Audio-7B-Instruct on the
same five ESC-50 event classes.  The original view
achieves 91.5\% accuracy. 
The majority vote ensemble achieves
94.5\%  accuracy.

We also evaluate learned aggregation using the official five
ESC-50 folds. Logistic regression achieves 95.5\% using all nine views.
Adding the transformed views yields eight additional correct
predictions, a gain of 4.0 percentage points over original-only
logistic regression. The largest recall gains are for Breath
(72.5\% to 82.5\%) and Sneeze (90.0\% to 97.5\%).
\begin{table}[t]
  \centering
  \caption{Qwen2-Audio accuracy on the five ESC-50 event classes.}
  \label{tab:qwen-aggregation}
  \begin{tabular}{@{}lrr@{}}
    \toprule
    Method & Views & Accuracy \\
    \midrule
    Original prediction & 1 & 91.5\% \\
    Majority vote (all views) & 9 & 94.5\% \\
    Logistic regression (all views) & 9 & \textbf{95.5\%} \\
    \bottomrule
  \end{tabular}
\end{table}

By analyzing the ablation of LR for Qwen, removing either the \(-6\) dB gain or the \(0.9\times\) time stretch reduced Qwen2-Audio LR accuracy from 95.5\% to 94.5\%. Removing the \(+2\) semitone pitch shift left Qwen2-Audio accuracy unchanged, whereas it caused the largest drop for SenseVoiceSmall.




\section{Discussion}
SoundVote improves event recognition when transformed views recover errors that other views miss, and the views that help most depend on the event class, the model, and how their predictions are combined.
\paragraph{Complementary views can recover errors.}
SoundVote tests whether a frozen audio model can improve an event decision
by voting over different presentations of the same recording. Our results
are consistent with an observation from vision: test-time augmentation can
correct some errors and introduce others, so net accuracy alone can hide
which decisions changed~\cite{shanmugam2021better}. In our audio experiments,
both pitch-shifted views are weaker than the original in isolation, yet
they improve accuracy when used together with it. The strength sweep also
shows that a stronger transformation can change more predictions without
improving the vote. For example, increasing the shift from $\pm2$ to
$\pm3$ semitones leaves ensemble accuracy at 70.5\%, even as predictions
change more often for the downward shift. These results motivate choosing
views according to the errors they recover together and the correct
predictions they preserve.
\paragraph{When to spend an extra inference call.}
The held-out SenseVoiceSmall router uses the original predicted label
to decide whether to query more views. It reaches 71.5\% accuracy at
0.48 seconds per clip, compared with 70.5\% at 1.09 seconds per clip for
the held-out fixed subset. The accuracy difference between those policies
is two clips. The larger difference is processing time. In this evaluation,
the router corrects 16 original errors without disrupting an originally
correct decision. It frequently selects extra views when the original
prediction is an unknown event or speech. A cost-matched comparison with
a simple rule that triggers on either label would help establish how
much benefit comes from learning separate subsets. Disagreement across
all eight transformations is only a weak indicator of original errors
(AUROC 0.599), so policies that use disagreement should account for the
cost of collecting those additional predictions and measure which errors
they still miss.
\paragraph{Scope of the evidence.}
The fixed nine-view ensemble improves accuracy on both ESC-50 and
Nonspeech7k, while accuracy on the same Nonspeech7k classes
remains below 50\%. This shows that the fixed recipe can help
under a different dataset's recording and labeling conditions. However, there is a ceiling to how much the proposed post-processing method can improve the predictions and beyond that we need to improve the models. 
The evaluations
cover five human non-speech classes with one reference label per clip.
They do not measure event localization, overlapping sounds, or false
alarms in continuous audio. 
\paragraph{Future work.}
One direction is to compare full-clip transformations with slicing
time windows and with combinations such as cropping followed by a mild
pitch shift. Short windows may preserve a transient event while reducing
competing context, and they can also omit the event entirely. Comparing
windows, individual transformations, and their compositions at the same
inference budget on held-out recordings would show when each is useful.
A second direction is to test the improvements on other tasks like speech emotion recognition (SER) and explore a
broader range of frozen audio models, including event taggers, classifiers,
and prompted audio-language models. Pitch, timing, and amplitude can
carry emotion cues, so SER requires checking whether a transformation
preserves the perceived emotion before using its prediction in a vote.
Evaluating the same candidate transformations across models and tasks
could reveal which view choices transfer and which require separate
development data.
A third direction is to test more complex transformations, such as
combining pitch shifts with frequency-selective filtering and noise addition, and to apply
the same evaluation framework in other domains, such as computer
vision. For example, we could test whether combining crops and
illumination changes helps recover image classification errors,
and whether those gains justify the additional inference cost.
\vspace{-0.2cm}
\section{Conclusion}
Small waveform changes can alter an audio foundation model’s prediction even when the recording contains the same event. We developed SoundVote to use this variation at inference time. SoundVote queries a frozen model on selected views of a recording and combines the predicted labels through voting or logistic regression. We evaluated how transformation strength, view selection, and aggregation affect recognition accuracy and inference cost.
On ESC-50, voting over the original and two pitch-shifted views raised SenseVoiceSmall accuracy from 63.5\% to 70.5\% at 0.68 seconds per clip. Voting over all nine views reached 73.0\% at 1.92 seconds per clip. Using those same nine views, logistic regression achieved held-out accuracies of 77.5\% for SenseVoiceSmall and 95.5\% for Qwen2-Audio. The fixed nine-view vote also raised SenseVoiceSmall accuracy from 40.74\% to 48.15\% on the five evaluated Nonspeech7k classes.
Our results show that transformed views can improve recognition by recovering different errors, and that the most useful views vary across event class. We will apply these findings to other audio tasks, such as speech emotion recognition, and investigate whether selecting and combining complementary views improves prediction reliability beyond audio.

\vspace{-0.2cm}

\bibliographystyle{ACM-Reference-Format}
\bibliography{ref} 

\end{document}